\documentstyle[bo99,epsfig,myastron]{article}

\def\grs{\mbox{GRS } 1915+105}

\def\irasnord{\mbox{IRAS } 19124+1106}
\def\irassud{\mbox{IRAS } 19132+1035}

\def\ss{\mbox{SS } 433}

\def\cm{\mbox{ cm}}

\def\mJy{\mbox{ mJy}}

\def\deg{^{\circ}}
\def\degp{{\rlap.}^{\circ}}
\def\amin{^\prime}

\def\asec{^{\prime \prime}}
\def\asecp{{\rlap.}^{\prime \prime}}

\def\heu{^{h}}

\def\hmin{^{m}}

\def\hsecp{{\rlap.}^{s}}

\def\douzecoabb{^{12} \mbox{CO}}
\def\treizecoabb{^{13} \mbox{CO}}
\def\hcoabb{\mbox{H}^{13}\mbox{CO}^{+}}

\def\csabb{\mbox{CS}}

\def\Ta{\mbox{T}_{A}^{*}}

\def\hdeuxromain{\mbox{H\,{\sc ii}}}

\def\ltsima{\; \buildrel < \over \sim \;}
\def\simlt{\lower.5ex\hbox{\ltsima}}            
\def\gtsima{\; \buildrel > \over \sim \;}
\def\simgt{\lower.5ex\hbox{\gtsima}}            

\title{ Search for interactions between ejections of GRS 1915+105
and its environment}

\author{ S. Chaty$^{1,2}$, L.F. Rodr\'{\i}guez$^3$ \& I.F. Mirabel$^2$ }
\affil{ $^1$Open University, Milton Keynes, UK; 
$^2$Service d'Astrophysique, CEA Saclay, France;
$^3$Instituto de Astronom\'{\i}a, UNAM, M\'exico}

\begin{document}

\maketitle

\begin{abstract}

To unravel the effect of likely interactions between the energetic 
ejections of the galactic superluminal source $\grs$ 
and its surrounding interstellar 
medium, we observed its environment. 
Two IRAS sources are symmetrically placed with respect to 
$\grs$, and are aligned with the
sub-arcsec ejections of this source.
 We analyzed these two sources $\irasnord$ and $\irassud$ through
 near-infrared, millimeter and centimeter wavelengths.
The evidence for these regions being interaction zones seems inconclusive.

\end{abstract}

\section{ Introduction }

The first known
galactic superluminal source $\grs$ is a highly energetic
and relativistically  
ejecting source. 
Consequently one wonders if there is an observable interaction
when the frequently ejected plasma clouds
collide at relativistic velocities with 
the interstellar medium, or heat molecular clouds
surrounding this source. 
Indeed, the ensemble of ejections of such a microquasar 
must have an effect on its environment, as it is the case
for the well-known ejecting source $\ss$.
Therefore, we undertook
a comprehensive study of the environment of $\grs$ at near-infrared,
mid-infrared, radio centimeter and millimeter wavelengths. 
Here some of the radio observations are described and discussed;
the reader can refer to Chaty et al. (2000, hereafter C00) 
for a complete description of the study.


\section{The observations: two axisymmetric sources}

The region surrounding $\grs$ was inspected and described by
Rodr\'{\i}guez and Mirabel (1998, hereafter RM98). 
This search was performed at $\lambda = 20$ cm, 
with the VLA (Very Large Array) of 
NRAO\footnotemark\footnotetext{The National Radio Astronomy Observatory 
is operated by Associated Universities, Inc., under cooperative agreement
with the USA National Science Foundation}, 
in C-configuration, giving a
 resolution of $15 \arcsec$. The resulting map is shown in Figure 1. 
They discovered that there were two axisymmetrically placed 
continuum radio sources, each
 located at $17 \arcmin$ from $\grs$, and coincident with IRAS sources.
Their positions are given in Table 1.
Furthermore, the position angle of these sources is 
$157 \pm 1 \deg$ from $\grs$, very similar to the one of the
well studied sub-arcsec radio-ejections from $\grs$ ($\sim 150 \deg$). 
In order to interpret the radio data, 
we remind here that the angle between the ejections
and the line of sight towards $\grs$ is $70 \deg$, that
the South component is approaching us, and the North component is receding 
(Mirabel and Rodr\'{\i}guez, 1994; Fender et al., 1999).

Although these two sources could be a chance alignment, 
the striking point-symmetric position of these two clouds suggests 
that they result from an association with the high-energy source $\grs$. 

\small

\begin{table}
\begin{flushleft}
\begin{center}
\begin{tabular}{|c|l|l|} \hline
{\em Source} & {\em J2000.0 coord.} & {\em gal. coord.} \\ \hline


GRS 1915+105 
& $\alpha = 19 \heu 15 \hmin 11 \hsecp 545$
& $l^{II} = 45 \degp 40 $                   \\

& $\delta = 10 \deg 56 \amin 44 \asecp 80$
& $b^{II} = -0 \degp 29 $                   \\ \hline


IRAS 19124+1106 
& $\alpha = 19 \heu 14 \hmin 45 \hsecp 77$
& $l^{II} = 45 \degp 54 $                              \\ 

& $\delta=11 \deg 12 \amin 06 \asecp 4$
& $b^{II} = -0 \degp 007 $  \\ \hline


IRAS 19132+1035        
& $\alpha = 19 \heu 15 \hmin 39 \hsecp 13$
& $l^{II} = 45 \degp 19 $    \\ 

& $\delta = 10 \deg 41 \amin 17 \asecp 1$
& $b^{II} = -0 \degp 44 $    \\

\hline
\end{tabular}
\end{center}
\end{flushleft}
\caption[]{\label{position} Positions of GRS 1915+105, IRAS 19124+1106 and 
IRAS 19132+1035.
These coordinates are the positions of peak signal 
obtained with the VLA-C 20 cm observations.
}
\end{table}

\normalsize


	\subsection{Centimeter wavelength observations}

High-resolution maps of the two continuum radio sources 
 have been obtained with the VLA (RM98). 
These maps are shown in the Figure 2. Concerning
the North lobe, the centimeter map shows that it resembles to a common
cometary H II region, but it also shows a shockwave structure to the South,
e.g. to the direction of $\grs$. For the South lobe, the centimeter map shows 
to the northwest a non-thermal jet, pointing along the direction of $\grs$.
The flux densities of this jet are $\sim\leq 1$, $2$ and $5 \mJy$ respectively
at $2, 6$ and $20 \cm$, showing a spectral index of $-0.8$.
 Furthermore, the South lobe shows a sharp
edge to the South, which could be either a bow shock, or an ionization
front in the $\hdeuxromain$ region.
The following discussion emphasizes these two striking 
features of the South lobe.



	\subsection{Millimeter wavelength observations}

We used the IRAM (Institut de Radio Astronomie 
Millim\'etrique) 30-m 
radio telescope, located on Pico Veleta, near Granada, Spain. 
The details of observations are described in C00,
and the results for $\irassud$ are shown in Figure 3.
The OFF position (position switching) was chosen at 
($\alpha = -500 \arcsec$, $\delta = -1200 \arcsec$) from $\grs$.
The main results are that i) the density profile of the cloud exhibits an
asymmetric velocity distribution, ii) the maximum of the profile
is closer to the counter-jet for high density tracers 
(compare e.g. $^{12}$CO 2-1 to the CS 2-1 transition),
and iii) there are two maxima in the $^{12}$CO 2-1 transition and only
one in the others. The other main result is that we detected a 4-$\sigma$ 
SiO 2-1 line, localized on the position of the counter-jet.

All these facts could indicate the presence of an interaction,
although they do not constitute a clear proof thereof.
Is there an association, 
or the two $\hdeuxromain$ regions are point-symmetric by chance?


\section{Discussion and Conclusion} \label{discussion}


There is a possibility that these two IRAS sources received energy from
$\grs$ through shocks initiated by plasma clouds ejected by $\grs$
and colliding with $\hdeuxromain$ regions, creating the non-thermal jet seen in 
the South lobe. There is also a possibility
that the relativistic ejecta have induced star formation, 
and this could have created the non-thermal
jet as a Herbig-Haro-like feature.
However, we consider this last possibility as unlikely because of
the timescale of the different phenomena.

The other possibility is that these two IRAS sources have
nothing to do with $\grs$: 
the alignment could be a background coincidence. It is worthwhile
to remember that there are two point-symmetric sources, and furthermore 
the IRAS fluxes and the molecular lines show that the two IRAS sources
 are in our Galaxy (RM98). 
This decreases the probability of a background coincidence.

Although our observations spanned in a large range of 
wavelengths (C00), and there are some striking facts, 
we can not clearly 
prove any association between $\grs$ and the two IRAS sources.


\begin{acknowledgements}


S.C. acknowledges support from grant F/00-180/A from the Leverhulme Trust,
and is grateful to C.A. Haswell for improving the language of the manuscript.

\end{acknowledgements}

\twocolumn
\small

\begin{figure}
\vspace*{-1.4cm}
\centerline{\psfig{file=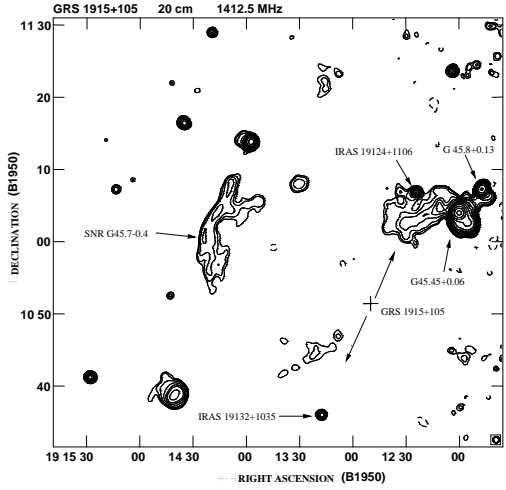,angle=0.,width=6.8cm}}
\vspace*{-1cm}
\caption[]{\small Map of the surroundings of $\grs$ (VLA-C, $\lambda = 20 \cm$, 
 half power contour of the beam shown 
in the bottom right corner). The arrows around $\grs$ indicate the position angle of the sub-arcsec relativistic ejecta. \label{grs-environs}}
\end{figure}

\begin{figure}
\vspace*{-0.7cm}
\centerline{\psfig{file=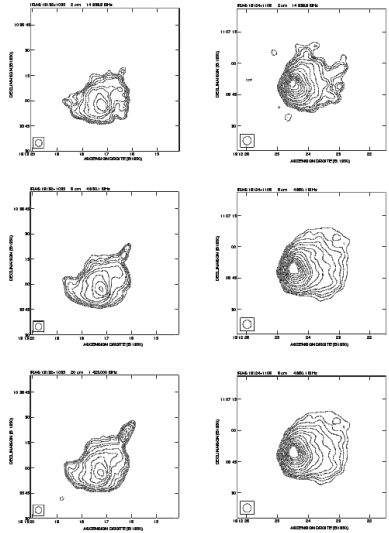,angle=0.,width=5.5cm}}
\caption[]{\small Maps of the two continuum radio sources $\irassud$ (left)
and $\irasnord$ (right), (VLA-D $\lambda = 2 \cm$ (top), C $6 \cm$ (middle) and 
B $20 \cm$ (bottom), half power 
contour of the beam shown in the bottom right corner. \label{les2lobes}
} 
\end{figure}

\begin{figure}
\vspace*{-.5cm}
\centerline{\psfig{file=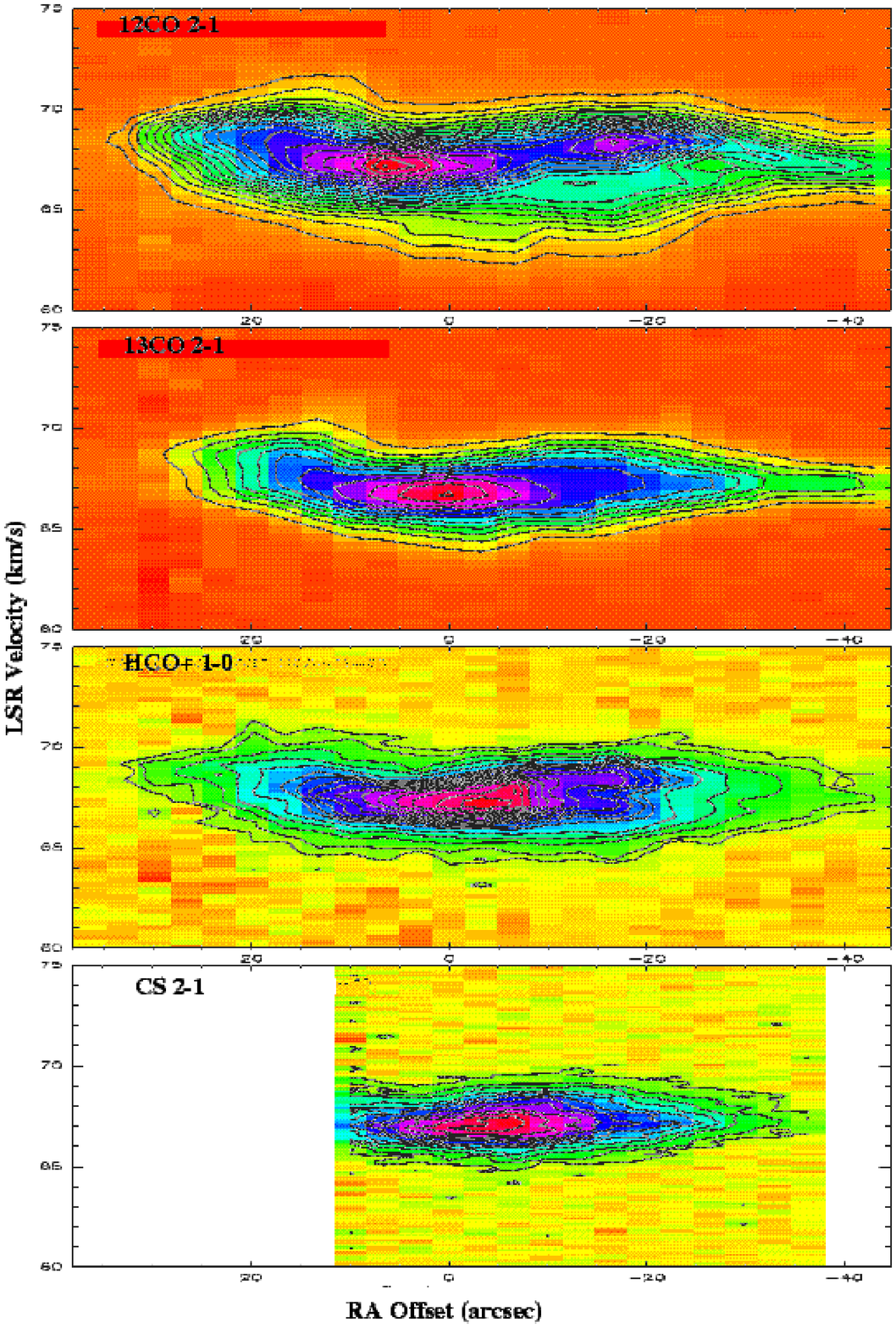,angle=0.,width=6.cm}}
\caption[]{\small Observations of IRAS 19132+1035.
Offsets are relative to the position of maximum radio emission
observed at the VLA.
 The transitions are, 
from top to bottom of Figure $\douzecoabb$, $\treizecoabb$, $\hcoabb$ and
 $\csabb$. The black contours are antennae iso-temperature, equal
respectively from top to bottom: 
$\douzecoabb$: $\Ta = -1$, and from $1$ to $20 K$ 
separated by an interval of $1 K$\-; 
$\treizecoabb$: $\Ta = -1$, and from $1$ to 11 K 
separated by an interval of 1 K\-;
$\hcoabb$: $\Ta = -1$, and from $0.2$ to $2.1 K$ 
separated by an interval of $0.1 K$\-; 
$\csabb$: $\Ta = -1$, and from $0.2$ to $2.2 K$ 
separated by an interval of $0.2 K$. The counter-jet seen in the VLA centimeter
continuum is located at $\sim -18 \asec$. \label{lobesud}
} 
\end{figure}



\end{document}